\documentclass{article}

\usepackage{graphicx,psfrag,epsf}
\usepackage{enumerate}
\usepackage{mathtools}
\usepackage{booktabs}
\usepackage{color}
\usepackage[english]{babel} 
\usepackage[protrusion=true,expansion=true]{microtype} 
\usepackage{amsmath,amsfonts,amsthm}
\usepackage{dsfont}
\usepackage{amssymb}
\usepackage{chngcntr}
\usepackage[toc,page]{appendix}
\usepackage{textcomp}
\usepackage{url}

\usepackage{array}
\usepackage{booktabs} 
\usepackage{setspace}
\usepackage{amsmath}

\usepackage{soul}
\usepackage{multirow}
\usepackage{enumitem}
\usepackage{microtype} 
\usepackage[font = small,labelfont=bf,textfont=it]{subcaption}
\usepackage{xcolor}
\usepackage{tabularx}
\usepackage[font = small,labelfont=bf,textfont=it]{caption} 
\usepackage{footnote}
\usepackage{algorithm}
\usepackage[noend]{algpseudocode}
\usepackage{etoolbox}
\usepackage{tabularx}
\usepackage{authblk}
\usepackage{caption}
\usepackage{indentfirst}
\usepackage{enumitem}
\newlist{steps}{enumerate}{1}
\setlist[steps, 1]{label = Step \arabic*:}

\algblock{ParFor}{EndParFor}
\algnewcommand\algorithmicparfor{\textbf{for}}
\algnewcommand\algorithmicpardo{\textbf{do\ parallel}}
\algnewcommand\algorithmicendparfor{\textbf{end\ parallel\ for}}
\algrenewtext{ParFor}[1]{\algorithmicparfor\ #1\ \algorithmicpardo}
\algrenewtext{EndParFor}{\algorithmicendparfor}

\makeatletter
\def\BState{\State\hskip-\ALG@thistlm}
\newcommand{\distas}[1]{\mathbin{\overset{#1}{\kern\z@\sim}}}%

\newsavebox{\mybox}\newsavebox{\mysim}
\newcommand{\distras}[1]{%
  \savebox{\mybox}{\hbox{\kern3pt$\scriptstyle#1$\kern3pt}}%
  \savebox{\mysim}{\hbox{$\sim$}}%
  \mathbin{\overset{#1}{\kern\z@\resizebox{\wd\mybox}{\ht\mysim}{$\sim$}}}%
}

\newcommand{\be}{\begin{equation}}
\newcommand{\ee}{\end{equation}}
\newcommand{\bi}{\begin{itemize}}
\newcommand{\ei}{\end{itemize}}
\newcommand{\ben}{\begin{enumerate}}
\newcommand{\een}{\end{enumerate}}

\newcolumntype{K}[1]{\geq {\centering\arraybackslash}p{#1}}
\allowdisplaybreaks

\makeatother

\let\oldbibliography\thebibliography
\renewcommand{\thebibliography}[1]{\oldbibliography{#1}
\setlength{\itemsep}{0pt}} 

\newcommand{\blind}{1}

\patchcmd{\footnotemark}{\stepcounter{footnote}}{\refstepcounter{footnote}}{}{}
\usepackage{authblk}
\usepackage[margin=40pt,font=small,labelfont=bf]{caption}
\usepackage[vmargin=1in,hmargin=1.5in]{geometry}
\usepackage{cleveref}

\begin{document}

\author[1]{Xiaojun Zheng}
\affil[1]{Duke University, \texttt{xz264@duke.edu}}

\author[2]{Simon Mak}
\affil[2]{Duke University, \texttt{sm769@duke.edu}}

\author[3]{Yao Xie}
\affil[3]{Georgia Institute of Technology, \texttt{yao.xie@isye.gatech.edu}}

\def\spacingset#1{\renewcommand{\baselinestretch}%
{#1}\small\normalsize} \spacingset{1}

\if1\blind
{
  \title{\bf Online High-Dimensional Change-Point Detection using Topological Data Analysis}
  \small
  \maketitle
} \fi

\if0\blind
{
  \bigskip
  \bigskip
  \bigskip
  \begin{center}
    {\LARGE\bf Online High-Dimensional Change-Point Detection using Topological Data Analysis}
\end{center}

  \medskip
} \fi



\begin{abstract}

Topological Data Analysis (TDA) is a rapidly growing field, which studies methods for learning underlying topological structures present in complex data representations. TDA methods have found recent success in extracting useful geometric structures for a wide range of applications, including protein classification, neuroscience, and time-series analysis. However, in many such applications, one is also interested in sequentially detecting changes in this topological structure. We propose a new method called Persistence Diagram based Change-Point (PD-CP), which tackles this problem by integrating the widely-used persistence diagrams in TDA with recent developments in nonparametric change-point detection. The key novelty in PD-CP is that it leverages the distribution of points on persistence diagrams for online detection of topological changes. We demonstrate the effectiveness of PD-CP in an application to solar flare monitoring.

\end{abstract}



\section{Introduction} \label{sec:intro}

Topological Data Analysis (TDA) is a thriving field that uses topological tools to study complex datasets' shapes and structures. In the modern era of big data, TDA provides an attractive framework for extracting low-dimensional geometric structures from such data, which are oftentimes high-dimensional and noisy. TDA methods have found recent success in a wide range of applications, including protein structure \cite{protein}, time-series data \cite{ts}, and neuroscience \cite{sizemore}. 

Despite such developments, there has been little work on integrating topological structure for change-point detection. Here, change-point detection refers to the detection of a possible change in the probability distribution of a stochastic process or time series. The need for change-point detection arises in many areas, from solar imaging to neuroscience, and the data in such applications exhibit topological structure as well. A recent work, \cite{ts_cp}, proposes an approach for time series data, by converting such data to a sequence of Betti numbers prior to estimating change-points. However, Betti numbers can only capture the number of features at pre-specified scales, while a persistence diagram (introduced in Section 2) preserves more topological information from the data. Persistence diagrams also enjoys a stability property \cite{stability}, which provides robustness under small perturbations of the data. This robustness is crucial for change-point detection, since a model needs to learn topological structure from noisy data prior to a change, before such structure can be used for identifying potential changes. 



We propose a new method called Persistence Diagram based Change-Point (PD-CP), which integrates persistence diagrams and a recently proposed non-parametric change-point detection approach in \cite{yao_cp}. Section 2 provides background on persistent homology. Section 3 outlines the PD-CP methodology. Section 4 demonstrates the effectiveness of this method on a solar flare monitoring application.


\section{Background} 

We first review a primary tool in TDA called \textit{persistent homology}, which extracts topological features (e.g., connected components, holes, and their higher-dimensional analogs) from point cloud data. Further details can be found in \cite{ghristbarcodes} and \cite{persistenthomology}.

\begin{figure}[!t]
    \centering
    \includegraphics[scale=0.35]{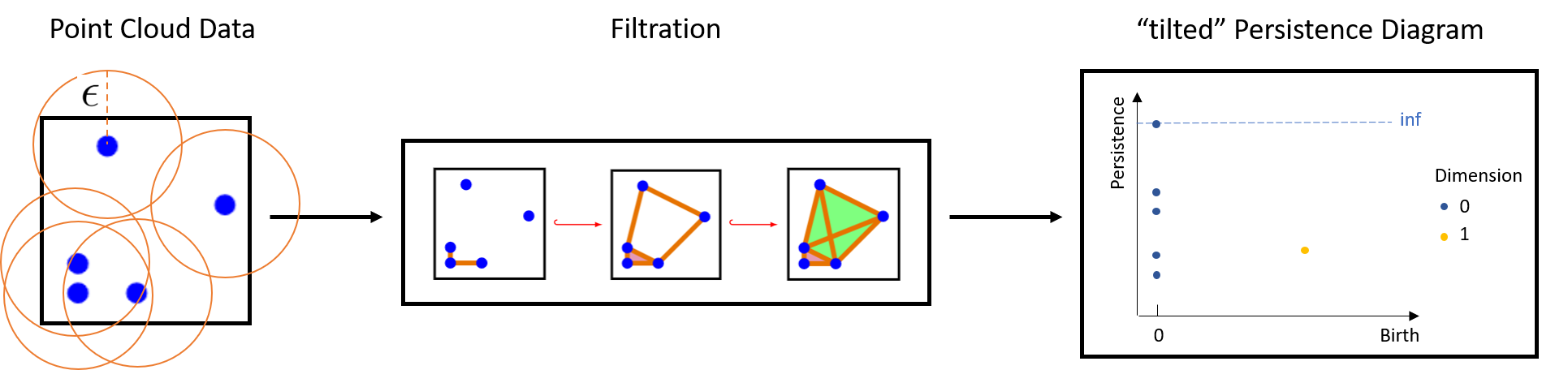}
    \caption{An illustration of the persistent homology pipeline, from point cloud data to a filtration of simplicial complexes to a (tilted) persistence diagram. The Rips complex with radius $\epsilon$ in the left plot corresponds to the second simplicial complex in the filtration.}
    \label{fig:tda}
\end{figure}

For a given point cloud dataset, persistent homology represents this point cloud as a \textit{simplicial complex}, defined as a set of vertices, edges, triangles, and their $n$-dimensional counterparts. A common simplicial complex built from point cloud data is the so-called Rips complex, which depends on a single scale parameter $\epsilon$.
At any $\epsilon > 0$, the Rips complex contains all edges between any two points whose distance is at most $\epsilon$, and contains triangular faces for any three points whose pairwise distance is at most $\epsilon$. Figure \ref{fig:tda} illustrates this for a toy dataset, adapted from \cite{fig_cite}. Clearly, a single scale parameter $\epsilon$ cannot capture all geometric structures of the data. Thus a sequence of scale parameters is used to build a \textit{filtration} of simplicial complexes. This filtration provides a means for extracting key topological structures from the data, such as the number of zero-dimensional holes (connected components) and one-dimensional holes. 

Under this framework, a topological feature appears in the filtration at some $\epsilon$ and disappears at some $\epsilon' > \epsilon$. The pair $(\epsilon, \epsilon')$ then gives the so-called \textit{persistence} of the feature, with $\epsilon$ and $\epsilon'$ being its \textit{birth} and \textit{death}, respectively. A large topological feature in the point cloud data would have long persistence, whereas a small or noisy topological feature would have short persistence. The collection of features can then be summarized by a {\it barcode}, where each bar has endpoints that correspond to the birth (i.e., $\epsilon$) and death (i.e., $\epsilon'$) of a feature. The information in a barcode can also be captured in a ``tilted'' \textit{persistence diagram}, in which a bar (representing a feature) is plotted as a point $(a,b)$, with $a = \epsilon$ is its birth time and $b = \epsilon' - \epsilon$ is its persistence time. Figure \ref{fig:tda} illustrates this tilted persistence diagram for the earlier toy dataset. This is slightly different from standard persistence diagrams, where $a$ and $b$ are taken to be the birth and death times, respectively.


While the above pipeline is presented for point cloud data, there are analogous approaches in the literature for building simplicial complexes and filtrations of more complex data types, e.g., time series \cite{ts} and image data \cite{image_ls}. The detection methodology presented next, which relies on the extracted persistence diagrams, can therefore be applied for these data types as well (see Section 4 for a solar flare monitoring application).

\section{Persistence Diagram based Change-point Detection}


Next, we introduce the proposed Persistence Diagram based Change-Point (PD-CP) method, which utilizes the extracted persistence diagrams over time for online detection of abrupt topological changes. We assume that the persistence diagrams outlined earlier are obtained for the data at each time $t = 1, \cdots, T$. PD-CP involves two key steps: (i) a histogram representation is constructed for each persistence diagram over time, and (ii) an online non-parametric hypothesis test is performed on these histograms to detect abrupt changes sequentially.

\begin{figure}[!t]
    \centering
    \begin{tabular}{cc}
         \includegraphics[width=.6\textwidth]{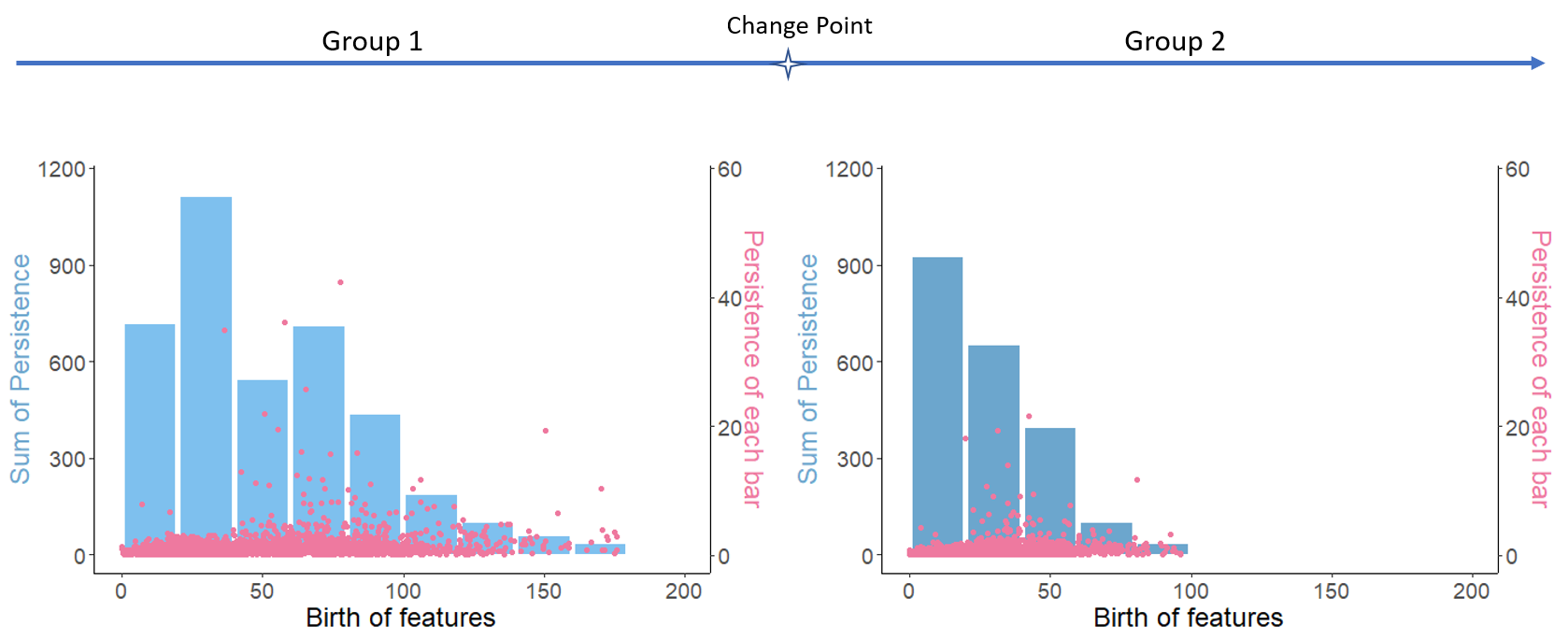}  & \includegraphics[width=.33\textwidth]{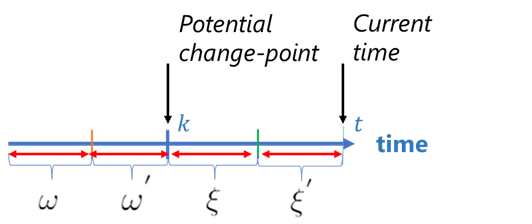} \\
        (a)  & (b)
     \end{tabular}
    \caption{(a) Histograms for the persistence diagram of a pre-change and post-change solar flare image. (b) Visualizing the intervals used for the weighted $\ell_2$ detection statistic.}
    \label{fig:hist_cp}
\end{figure}

Consider the first step (i). To construct a histogram that captures topological information from a persistence diagram, we split the domain for birth times into $M$ different bins, then sum up the persistence of features within each bin. This histogram binning serves two purposes: it provides a robust way for reducing noise in the persistent diagram data, and allows us to leverage recent developments in empirical distribution based change-point methods. Figure \ref{fig:hist_cp}(a) visualizes this construction. The breakpoints for these bins (denoted as $b_1, \cdots, b_M$) are trained using the ``pre-change'' persistence diagrams (i.e., the diagrams before the abrupt change) and are kept the same throughout the procedure. Figure \ref{fig:hist_cp}(a) (left) shows this for a solar flare image (see Section 4) prior to an abrupt change. After a change-point, the ``post-change'' persistence diagrams are binned using the same breakpoints. These post-change histograms are then expected to be significantly different from the pre-change histograms. Figure \ref{fig:hist_cp}(a) (right) shows the histogram for a post-change solar flare image.


Consider the second step (ii). To detect differences between pre-change and post-change histograms, we make use of a non-parametric detection statistic recently proposed in \cite{yao_cp}, which uses a weighted $\ell_2$ divergence between the two histograms (representing empirical distributions) to  detect changes sequentially. Our approach is as follows.  At a given time $t$, we search for all possible change-points at time $k < t$. To investigate whether time $k$ is a change-point, we will consider four consecutive time intervals (see Figure \ref{fig:hist_cp}(b)): the first two intervals are immediately before time $k$ and the last two are immediately after $k$, with all intervals having the same length. We call the former intervals ``group 1'' and the latter ``group 2'', representing potentially pre-change and post-change times. Let $\omega_{t,k}, \omega'_{t,k} \in \mathbb{R}^M$ be the empirical distributions of persistence diagrams from the two intervals in group 1 (binned using breakpoints $b_1, \cdots, b_M$), and $\xi_{t,k}, \xi_{t,k}' \in \mathbb{R}^M$ be the empirical distributions of observations from the two intervals in group 2. Let $\Sigma = \text{Diag}\{\sigma_1. \cdots, \sigma_M\}$ be a weight matrix, where $\sigma_m \geq 0, m=1, \cdots, M$. The weighted $\ell_2$ statistic can then defined as
\[\chi_{t,k} = (\omega_{t,k} -\xi_{t,k})^T \Sigma (\omega'_{t,k}-\xi'_{t,k}).\]
A larger value of $\chi_{t,k}$ gives greater evidence of a change-point at time $k$, using data up to time $t$. 


An online detection procedure is then given by the stopping time:
\[ \mathcal{T} = \inf\{t: \chi_t^{\max} \geq b\}, \quad \chi_t^{\max} = \max_{0 \leq k \leq t} \chi_{k,t},\]
where $b$ is a pre-specified threshold parameter. Here, $\mathcal{T}$ is the time at which the procedure raises an alarm indicating a change-point has occurred before time $t$, by taking the maximum statistic $\chi_t^{max}$ over all possible change-points $k < t$. The threshold $b$ is typically set by controlling the false alarm rate to be below a certain pre-specified level (see \cite{yao_cp}).



\section{Detecting Solar Flare Changes}


Solar flares are sudden flashes of brightness on the sun. Such flares are closely related to geomagnetic storms, which can cause large-scale power-grid failures. In recent years \cite{Sun2021}, the sun has entered a phase of intense activity, which makes monitoring solar flares an important task \cite{yao_solar}. However, these flashes are hardly visible and can be missed by a baseline detection statistic, thus making monitoring a difficult task. We demonstrate the effectiveness of PD-CP in detecting changes in a sequence of solar images ($232 \times 292$ pixels) at times $t = 1, \cdots, T=300$; this data is obtained from the Solar Dynamics Observatory\footnote{See \url{https://sdo.gsfc.nasa.gov/mission/instruments.php}.} at NASA.



To begin, however, we would need to define an appropriate filtration for capturing topological features in images. We make use of the \textit{lower star filtration}, which have been used for topological analysis of images \cite{image_ls}. For a real-valued function $f: \mathcal{X} \rightarrow \mathbb{R}$, define the sublevel set of $f$ as:
\begin{equation}
X(\epsilon) = \{x \in \mathcal{X} | f(x) \leq \epsilon\}.
\label{eq:sublevel}
\end{equation}
For a finite set of $\epsilon_1, \epsilon_2, \cdots, \epsilon_n > 0$, a \textit{sublevel set filtration} of $X$ is then defined as the sequence of simplicial complexes $X_1 \subset \cdots \subset X_n$, where $X_i = X(\epsilon_i)$, $i = 1, \cdots, n$. The filtration provides a characterization of topological structure on $f$.

Sublevel set filtrations provide a natural persistent homology for images, by viewing an image as a function mapping each pixel location to its intensity value. Considering the image pixels as vertices on a grid, we first triangulate this grid by placing an edge between two points that are horizontally, vertically, or diagonally adjacent, and a triangular face for any three adjacent points forming a triangle. Using image intensity values as the response for $f$ in \eqref{eq:sublevel}, the sublevel set filtration $X_1 \subset \cdots \subset X_n$ then forms a sequence of simplicial complexes.

\begin{figure}[!t]
     \centering
     \begin{tabular}{cc}
         \includegraphics[width=.4\textwidth]{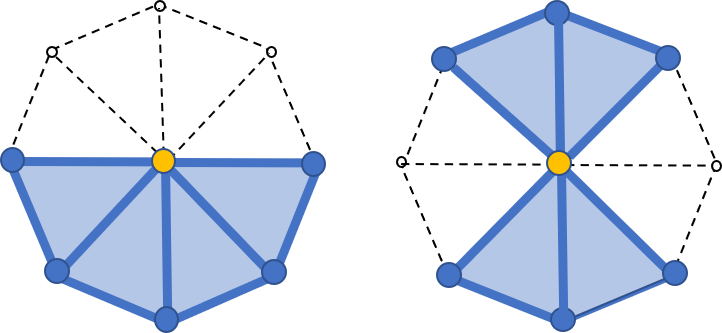}  & \includegraphics[width=.35\textwidth]{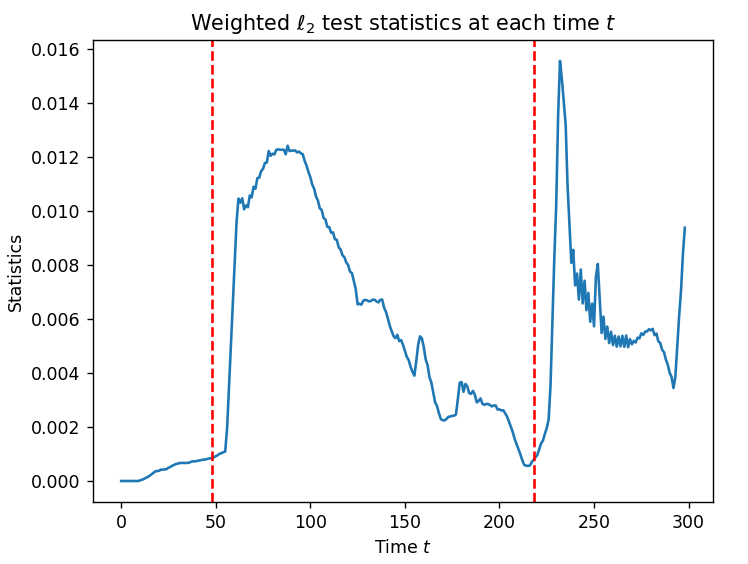} \\
        (a)  & (b)
     \end{tabular}
     \caption{(a) Visualizing a regular point (left) and a saddle point (right) in the lower star filtration. (b) The detection statistic $\chi_t^{max}$ at each time $t$, with red dashed lines indicating the true change-points.}
        \label{fig:three graphs}
\end{figure}

When a new vertex is added in the sublevel set, the topological change depends on whether the vertex is a maximum, minimum, regular, or a saddle of the function. Figure \ref{fig:three graphs}(a) visualizes a regular point and saddle point (in yellow), and the edges and faces in the sublevel sets (in blue). The topological features do not change after introducing a regular point, but the number of connected components decreases by one after introducing a saddle point. This filtration provides a means for extract image topological features as persistence diagrams.

\begin{figure}[!t]
    \centering
    \begin{tabular}{cc}
         \includegraphics[width=.45\textwidth]{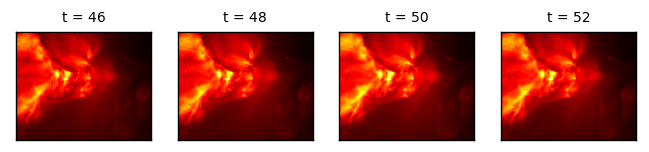}  & \includegraphics[width=.45\textwidth]{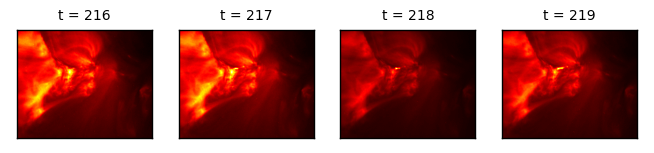} \\
    \end{tabular}
    \caption{Snapshots of the solar flare at two change-points $t^*_1=50$ and $t^*_2 = 218$.}
    \label{example_images}
\end{figure}

We then integrate this sublevel set filtration within the detection framework in Section 3, to detect topological changes for the aforementioned solar flare problem. The histogram breakpoints $b_1, \cdots, b_M$ are chosen such that there is (roughly) an equal sum of persistences within each bin for the first solar flare image. Figure \ref{fig:three graphs}(b) shows the detection statistic $\chi_t^{\max}$ as a function of time $t$, using $M=10$ bins for histograms. We see two sudden increases in the statistic $\chi_t^{\max}$, one after time $t^*_1=50$, and another after $t^*_2=218$. These are dotted in red in the figure, and suggests a change-point in topological structure. To investigate further, Figure \ref{example_images} shows snapshots of the solar flare immediately before and after $t_1^*$ and $t_2^*$. For both times, we see a clear change-point in the images: at $t^*_1 = 50$, the flare bursts become more pronounced and bright, whereas at $t^*_2 = 218$, certain flares become noticeably more subtle and subdued. The proposed PD-CP approach appears to nicely capture this change with little detection delay, given an appropriately set threshold. 




We also note that the PD-CP is quite computationally efficient in this experiment. Using the Python package Ripser \cite{ripser}, the computation time for building the lower star filtration of both connected components and holes on all $T=300$ images is approximately $90$ seconds on a standard desktop computer. Given this filtration, the detection statistic $\chi_t^{\max}$ can be then evaluated with minimal additional computation, which allows for efficient online detection.

\section{Conclusion}
We have proposed a Persistence Diagram based Change-Point (PD-CP) method, which integrating the persistence diagrams from TDA with a nonparametric change-point detection approach. The idea is to first learn topological structure via persistence diagrams, and use a weighted $\ell_2$ divergence on a histogram representation of these diagrams to sequentially detect topological change. There are several interesting directions of future research. First, we are aiming to utilize the persistence diagrams on both connected components and holes, and integrate this within PD-CP. Second, we are exploring a more localized detection approach, which can better identify local changes (e.g., local translation / rotation shifts) in images.

\bibliographystyle{unsrt}
{\footnotesize \bibliography{iclr2021_workshop}}

\end{document}